\documentclass{emulateapj}
\def\ltsima{$\; \buildrel < \over \sim \;$}
\def\simlt{\lower.5ex\hbox{\ltsima}}
\def\gtsima{$\; \buildrel > \over \sim \;$}
\def\simgt{\lower.5ex\hbox{\gtsima}}

\def\gsim{\mathrel{\raise0.35ex\hbox{$\scriptstyle >$}\kern-0.6em
\lower0.40ex\hbox{{$\scriptstyle \sim$}}}}
\def\lsim{\mathrel{\raise0.35ex\hbox{$\scriptstyle <$}\kern-0.6em
\lower0.40ex\hbox{{$\scriptstyle \sim$}}}}

\def\gs{\mathrel{\raise0.35ex\hbox{$\scriptstyle >$}\kern-0.6em
\lower0.40ex\hbox{{$\scriptstyle \sim$}}}}
\def\ls{\mathrel{\raise0.35ex\hbox{$\scriptstyle <$}\kern-0.6em
\lower0.40ex\hbox{{$\scriptstyle \sim$}}}}



\slugcomment{{\sc accepted in  ApJ:  Aug17, 2008}  }

\begin{document}

\title{Do submillimeter galaxies really trace the most massive dark matter halos? Discovery of a high-$z$ cluster in a highly active phase of evolution.  }
\author{S.\ C.\ Chapman\altaffilmark{1,2,3},  
A.\ Blain\altaffilmark{4}, R.\ Ibata\altaffilmark{5}, 
R.\ J.\ Ivison\altaffilmark{6,7}, I.\ Smail\altaffilmark{8}, G.\ Morrison\altaffilmark{9,10}
}
\altaffiltext{1}{Institute of Astronomy, Madingley Road, Cambridge, CB3 0HA, U.K.}
\altaffiltext{2}{Canadian Space Agency Fellow}
\altaffiltext{3}{Department of Physics and Astronomy, University of Victoria, Victoria, B.C., V8P 1A1, Canada}
\altaffiltext{4}{California Institute of Technology, Pasadena, CA\,91125}
\altaffiltext{5}{Observatoire de Strasbourg, 11, rue de l'Universit\'e, F-67000, Strasbourg, France}
\altaffiltext{6}{UK Astronomy Technology Centre, Royal Observatory, Blackford Hill, 
Edinburgh EH9 3HJ, UK.}
\altaffiltext{7}{Institute for Astronomy, University of Edinburgh,     Blackford Hill, Edinburgh EH9 3HJ, UK.}
\altaffiltext{8}{Institute for Computational Cosmology, Durham University, South Road, Durham DH1 3LE, UK.}
\altaffiltext{9}{Institute for Astronomy, University of Hawaii, Honolulu, HI, 96822, USA}
\altaffiltext{10}{Canada-France-Hawaii Telescope, Kamuela, HI, 96743, USA}

\begin{abstract}
We present detailed observations of a $z\sim1.99$ cluster of submillimeter 
galaxies (SMGs), discovered as the strongest
redshift spike in our entire survey of $\sim$100 SMGs across 800 square arcmin.
% survey over 8 fields on the sky -- 
It is the largest blank-field SMG concentration currently known %from all surveys 
and has $<0.01$\% 
chance of being drawn from the underlying selection function for SMGs. 
We have compared UV observations of galaxies at this redshift, where we
find a much less dramatic overdensity, having an 11\% chance of being drawn from its selection function.
We use this $z\sim1.99$ overdensity to compare the biasing of UV- and submm-selected galaxies, and test
whether SMGs could reside in less overdense environments, with their apparent
clustering signal being dominated by highly active merger periods in modest mass structures.
We discuss the probable
mechanisms for the apparently different bias we see at the two wavelengths.
%, and argue for a modest mass
%galaxy cluster seen at a particularly formative period in its evolution.
This impressively active formation 
 phase in a low mass cluster is not something seen in simulations, although we propose a toy model
 using {\it merger bias} which could account for the bias seen in the SMGs.
While enhanced buildup of stellar mass appears characteristic of other high$-z$ galaxy clusters, 
neither the UV- nor submm-galaxies in this structure exhibit larger stellar masses than their field galaxy counterparts (although the excess of SMGs in the structure represents a larger volume-averaged stellar mass than the field).
Our findings have strong implications for future 
surveys for high-z galaxies at long wavelengths such as SCUBA2 and {\it Herschel}.
%\footnote{SCUBA2, Herschel, CCAT, ALMA}.
We suggest that since these surveys will select galaxies during their episodes of peak starbursts,
they could probe a much wider range of environments than
just the progenitors of rich clusters, revealing more completely  
the key events and stages in galaxy formation and assembly.

%surveys for high$-z$ clustered galaxies at long wavelengths (SCUBA2, Herschel), where one is biased to selecting the most luminous episode in a given galaxy regardless of its timescale or environment.
%
\end{abstract}

\keywords{galaxies: optically faint radio 
 --- galaxies: evolution}

\section{Introduction}
\label{txt:intro}

Submillimeter galaxies (SMGs) have provided an efficient probe of ultraluminous ($>10^{12}$~L$_\odot$) star formation activity in the distant Universe (e.g, Smail, Ivison \& Blain 1997; Blain et al.\ 2002; Borys et al.\ 2003; Coppin et al.\ 2008; Knudsen et al.\ 2008), with the bright submm detection (rest frame far-infrared -- FIR) providing an unambiguous signal of large dust masses heated predominantly by young stars rather than AGN (e.g., Almaini et al.\ 1999; Fabian et al.\ 2000;
Alexander et al.\ 2005; Menendez-Delmetre et al.\ 2007; Pope et al.\ 2008).
Piecing together the detailed properties of SMGs has been difficult because of the inherent dust obscuration to the regions of the largest bolometric output.
High spatial resolution radio observations using the MERLIN interferometer (Chapman et al.\ 2004a; Biggs \& Ivison 2008) provided early indications that the dust and gas in SMGs was more compact than typical galaxies, yet still resolved on scales three times larger than in local ULIRGs. Once verified directly through high resolution CO gas studies of SMGs (Tacconi et al.\ 2006, 2008) this was used to argue that SMGs are scaled up ULIRGs. 
This radio and CO mapping of SMGs % high resolution radio (Chapman et al.\ 2004)
has suggested that the locus of far-IR emission is often offset from UV/optical emission, consistent with findings (Chapman et al.\ 2005 -- C05) that even dust-corrected UV estimates  of star formation rates in SMGs under-predict the true star formation rates (SFRs) by factors $>100\times$ on average.
%it has only been through
CO linewidths for a sizable sample of SMGs (Frayer et al.\ 1998, Neri et al.\ 2003, Greve et al.\ 2005) have suggested large dynamical masses, $>10^{11} M_\odot$, though not nearly as large as the halos in which they reside based on their clustering properties (Blain et al.\ 2004).
%ZZZ - Need a reference here to some mass-r_0 link since it's the intro?
%
The large dynamical masses are associated with large ($1-2\times10^{11}$~M$_\odot$) stellar masses %as measured from the rest-frame $K$-band 
(Borys et al.\ 2005; Hainline 2008; Swinbank et al.\ 2008).
Finally, the gas masses and SFRs define an average gas exhaustion timescale $<$50~Myrs, suggesting a duty cycle of $\sim$10 for an underlying population of galaxies hosting the submm-luminous events.
These studies bring together a picture of SMGs as typically representing massive, gas rich galaxies, which ought to be building elliptical galaxies that are found preferentially  in  clusters at the present day (Swinbank et al.\ 2006, 2008).

The strongly 
clustered populations of red $K$-selected galaxies at $z=2-4$ (e.g., Daddi et al.\ 2003) and of luminous high-$z$ radio galaxies (HzRG -- Brand et al.\ 2005) suggest they are formed in the highest density perturbations at early epochs, implying that they are the progenitors of local massive early-type galaxies and $z\sim1$ extremely red objects (EROs) which themselves have similarly strong clustering (e.g., Brown et al.\ 2005).
HzRGs are believed to host massive black holes; thus may represent some of the most massive galaxies at these epochs (e.g., de Breuck et al.\ 2000) and could 
trace some of the highest density environments. 
The identiÞcation of modest excesses of companion galaxies around HzRGs in the 
optical/near-infrared (e.g., Rottgering et al.\ 1996), X-ray 
(Pentericci et al.\ 2002; Smail et al.\  2003a), and 
submm (Ivison et al. 2000; Stevens et al. 2003),
supports  their association with overdense structures.

Studies to date
have suggested that  the most FIR-luminous, distant galaxies  are also clustered very strongly
(Blain et al.\ 2004; Borys et al.\ 2004; Scott et al.\ 2002, 2006; van Kampen et al.\ 2005; Farrah et al.\ 2006;
Blake et al.\ 2006; Gilli et al.\ 2007). 
At face value this would indicate that they occupy the largest dark-matter halos ($\sim$10$^{13}$~M$_\odot$).
However selection of galaxies by the most FIR-luminous specimens may be prone to
finding the brief periods when systems of lower matter overdensity are rapidly evolving
(e.g.\ Scannapieco \& Thacker 2003; Furlanetto \& Kamionkowski 2006). Indeed burst timescales as short as 10~Myr have been estimated for some SMGs (e.g., Smail et al.\ 2003b).
Blain et al.\ (2004) calculated that, complex biases aside, SMGs have a clustering length which is consistent with a form of evolution ensuring their properties subsequently match the clustering length typical of evolved red galaxies at $z\sim1$ and finally of rich clusters of galaxies at the present epoch.

% Figure projection of z=2 structure
\begin{figure}
\centering
\includegraphics[width=7.9cm,angle=0]{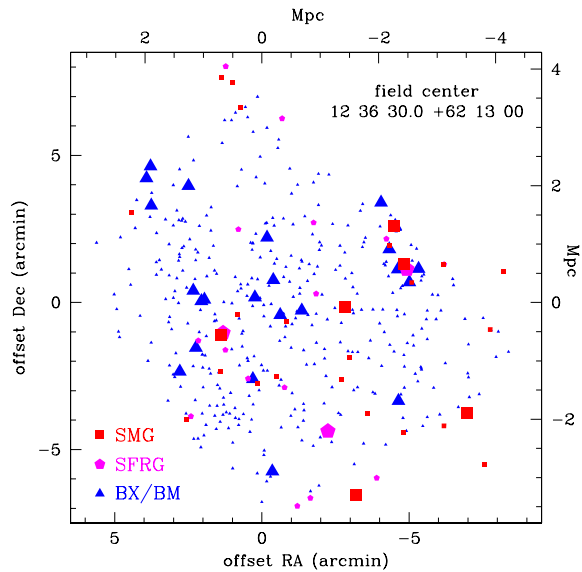}
\caption{\footnotesize
Positions of all UV-selected galaxies an SMGs/SFRGs in the survey region are shown as small symbols, while those lying in the $z=1.99$ structure are large symbols. Offset projected Mpc are shown on the axes for reference at $z=2$.
Galaxies in the $z=1.99$ structure is generally widely separated on the sky, with $>$Mpc separations between most SMGs/SFRGs except for two close ($\sim100$~kpc) SMG/SFRG pairs in two cases. 
}
\label{fig1}
\end{figure}

%
% Figure N(z)
%
\begin{figure}
\centering
\includegraphics[width=7.5cm,angle=0]{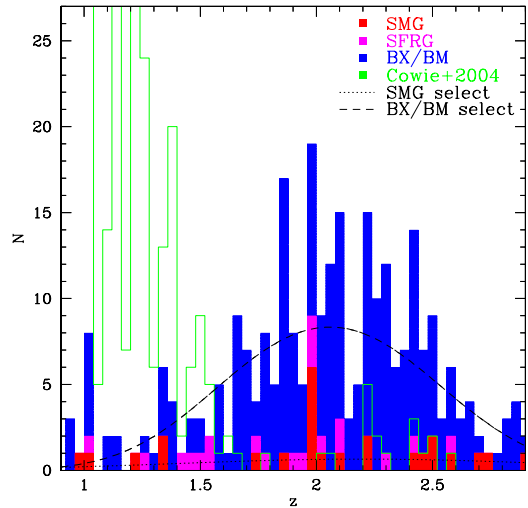}
\includegraphics[width=7.5cm,angle=0]{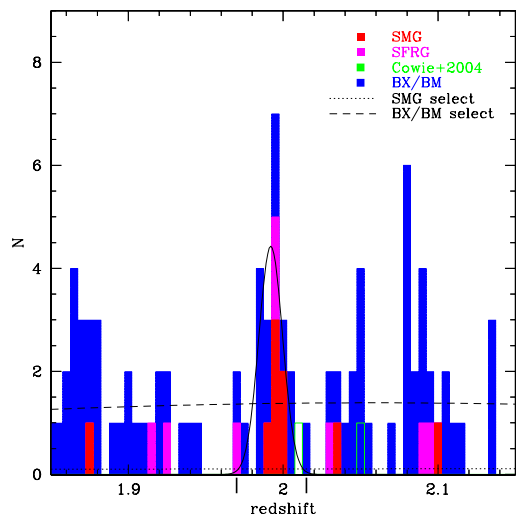}
\caption{\footnotesize
{\bf (top panel)} Redshift histograms in GOODS-N (bins of $\delta z= 0.02$).  
from the surveys of Reddy et al.\ (2006) in UV-selected BX/BM galaxies, {\it Team Keck Redshift Survey}  (Wirth et al.\ 2004), and Hawaii redshift survey (Cowie et al.\ 2004), the SMG surveys of Chapman et al.\ (2003, 2005), and the SFRG surveys of Chapman et al.\ (2004) and Casey et al.\ (2008).
The z=1.99 structure can be seen as the highest peak of galaxies in the $z\sim2$ region.
The selection functions for SMGs and BX/BMs are depicted normalized to the number of sources in each sample. The SFRGs are included in the SMG selection function (see text).
There are a few overlapping BX/BM, TKRS and SMGs/SFRGs which are effectively double-counted in the plot.
%{\bf (bottom panel)} A zoom in around the $z=1.99$ structure, where overlapping galaxies in the various samples have been removed, and the histogram now represents a cumulative histogram of all populations.
% (notably the two OFRGs at z=2.09), but none
Note that the BX and SMG/OFRG galaxy samples in the  $z=1.99$ structure are completely non-overlapping, and the structure could arguably be more significant than other BX galaxy peaks in the combined BX/SMG galaxy sample.
{\bf (lower panel)} A zoomed in view of the structure with smaller redshift bins ($\delta z= 0.005$).  
The spike galaxies listed in Table~1 are shown by the vertical bars.}
\label{fig2}
\end{figure}

%ZZZ - why pick 100 arcmin? I think 100 arcmin
%is about right to get one of these guys - since 100 arcmin is ~50 Mpc.

However, as pointed out by Blain et al.\ (2004),  this is inconsistent with the expected density of rich clusters. It  is very unlikely (0.05\% chance) that any field of order 10\arcmin$\times$10\arcmin\  includes the progenitor of a moderately rich cluster (volume density $\rho= 10^{-7}$Mpc$^{-3}$), even in the long pencil beam from $z=$1-3 of a radio-SMG survey. 
This apparent  paradox can only be resolved by
%this is inconsistent with the progenitor of a relatively rich cluster of galaxies intersecting any field out to a redshift $\sim3$ covering a solid angle of order 100 arcmin$^2$ on a side, an apparent paradox that can only be resolved by 
reducing the frequency of such associations by a factor of ten ($\rho=10^{-7}$Mpc$^{-3}$ observed for moderately rich clusters compared with  $10^{-6}$Mpc$^{-3}$ implied by SMGs).
Blain et al.\ concluded that either the dark matter halo masses are less extreme than inferred from a picture of simple 1-1 biasing at a given mass, or some other biasing effect must be at work.
% that SMGs appear to cluster as strongly
Adelberger et al.\ (2005a) presented an alternative explanation in the
inherent bias of the cluster calculation methodology, however statistical analyses of the SMG clustering through other methods is consistent with the Blain et al.\ (2004) calculation (e.g., Blake et al.\ 2006).
The alternative explanation, that SMGs often reside in less overdense environments (and that ULIRG
clustering could be dominated by the coordination of highly active periods across modest mass
environments), is
 tested here, using UV observations of the richest overdensity found  in
Blain et al.\ (2004) and Chapman et al.\ (2005).
%our own spectroscopic surveys.>> REFS.

In \S2 we describe the sample properties, with \S2.1 assessing the various selection functions, and a developing a statistically robust measure of the galaxy overdensity. 
In \S2.2 we analyze the properties of the galaxies lying in the overdense structure, including their star formation rates at a number of wavelengths, their luminosity function, and their angular positions on the sky.
In \S3 we discuss the theoretical expectations of the measured redshift space contrasts, propose an explanation for the differential bias seen at different wavelengths (\S3.1), and discuss the implications for future surveys at submm wavelengths (\S3.2).
We assume
a $\Lambda$-CDM cosmology (Spergel et al.\ 2003) with $\Omega_0=0.3$, $\Omega_\Lambda=0.7$ and
$H_0=70$\,km\,s$^{-1}$\,Mpc$^{-1}$, so that 1\,arcsec corresponds to
8.4\,kpc physical size at $z=2.0$.

\section{Characterizing the structure at UV- and submm- wavelengths}

Blain et al.\ (2004) highlighted an association of five SMGs in the GOODS-N field, all lying within 1200~km/s of $z=1.995$, representing the strongest association in their spectroscopic survey of 
73 SMGs. The GOODS-N field is one of the better sampled fields from the expanded C05 SMG spectroscopic survey and the ongoing spectroscopic surveys of the SHADES submm fields (e.g., Coppin et al.\ 2008, Blain et al.\ in prep), however associations of similar significance can be reasonably ruled out in these other survey fields (although the {\it association distribution function} N(n$_{SMG}$) is not yet well characterized). 
To further study this GOODS-N field SMG association and its properties at other wavelengths, we bring together an expanded sample of 
SMGs in this field, representing our primary sample of ultra-luminous star forming galaxies at high redshift. 
A population of submm-faint, radio-selected  galaxies (SFRGs) with similar inferred star formation rates to SMGs as evidenced by radio and 24$\mu$m luminosities
(Chapman et al.\ 2004, Casey et al.\ in prep.)
represent a secondary sample from which to study the cluster and test our hypotheses.
A map of source positions is shown in Fig.~1, depicting the
spatial distributions of all galaxies with known spectroscopic redshift $z>1$  in the field and highlighting the candidate $z=1.99$ structure. 
%Overview of radio/24/mm selection of active/star forming galaxies.

The GOODS-N sample of 30 SMGs comprises the spectroscopic redshift sample of  22 radio-identified SMGs in Chapman et al.\ (2005), and six additional radio-SMGs with spectroscopic redshifts presented in Pope et al.\ (2006, 2008), and two further radio-SMGs from Daddi et al.\ (2008). We also include 19 submm-faint {\it SFRGs} from Chapman et al.\ (2004) and Casey et al.\ (in prep.), and assemble a complete catalog of UV/optical selected galaxies with redshifts, combining the surveys of   Wirth et al.\ (2004), Cowie et al.\ (2004),  Reddy et al.\ (2006), with our own spectroscopic data. In Table~1 we list all known galaxies in the $z=1.99$ structure, including the several unpublished UV-selected galaxies from our surveys. 
We assemble the relevant archival multi-wavelength data in Table~1 which is discussed explicitly in this paper, as well as  new 1.4~GHz VLA radio measurements from the maps of  Morrison et al.\ (2008).
The redshift distributions are shown in Fig.~2.
The well studied GOODS-N region is covered by a superb multi-wavelength survey to great sensitivity in most wavebands, and so it is unlikely that further missing galaxy populations could skew our results significantly. However there is no spectroscopic sample available which effectively targets quiescent galaxies at $z\sim2$.

%Caption - include explicit SFR - L conversion.

\begin{figure}
\centering
\includegraphics[width=7.5cm,angle=0]{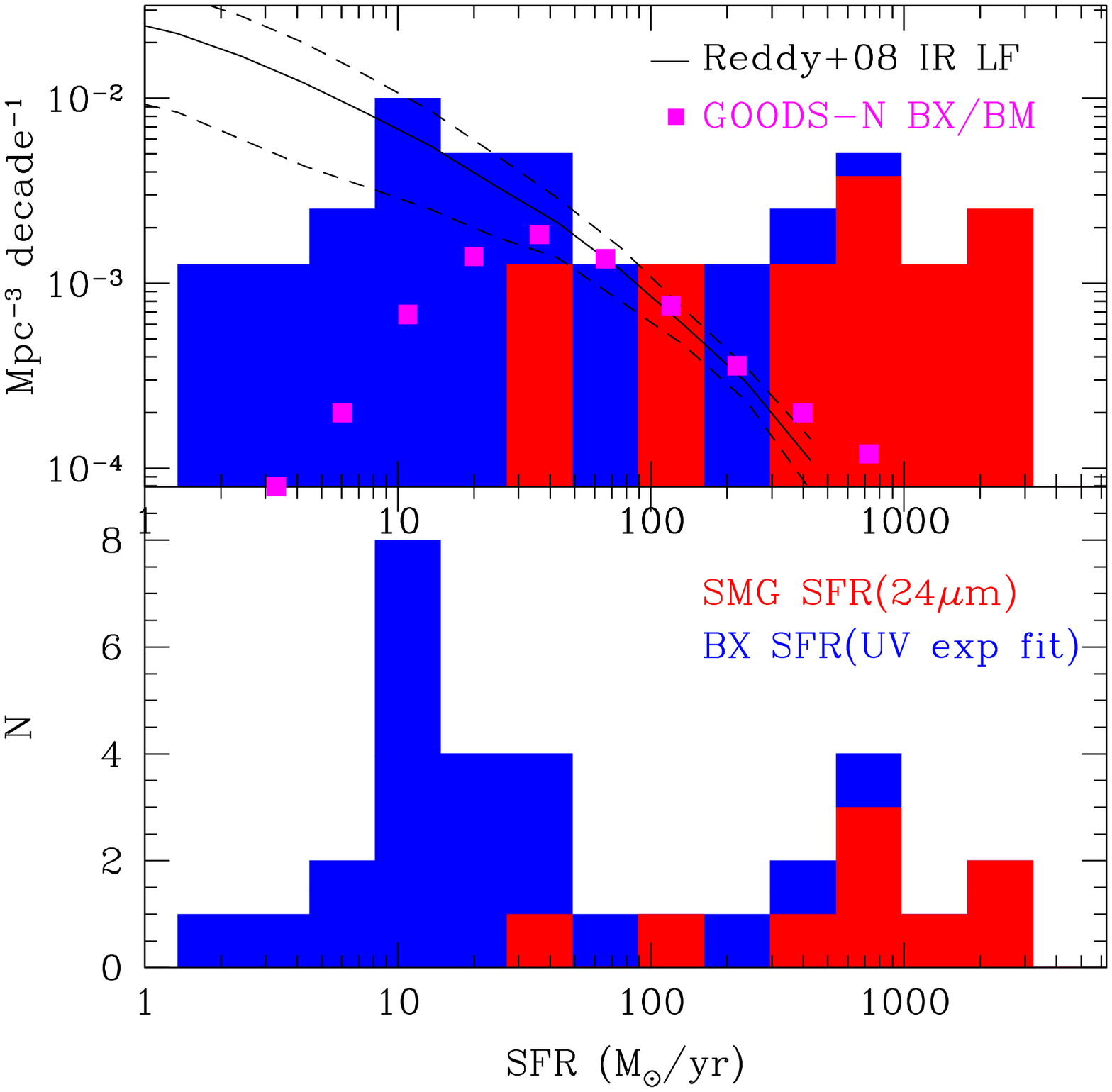}
\vskip-2.5cm
\includegraphics[width=7.5cm,angle=0]{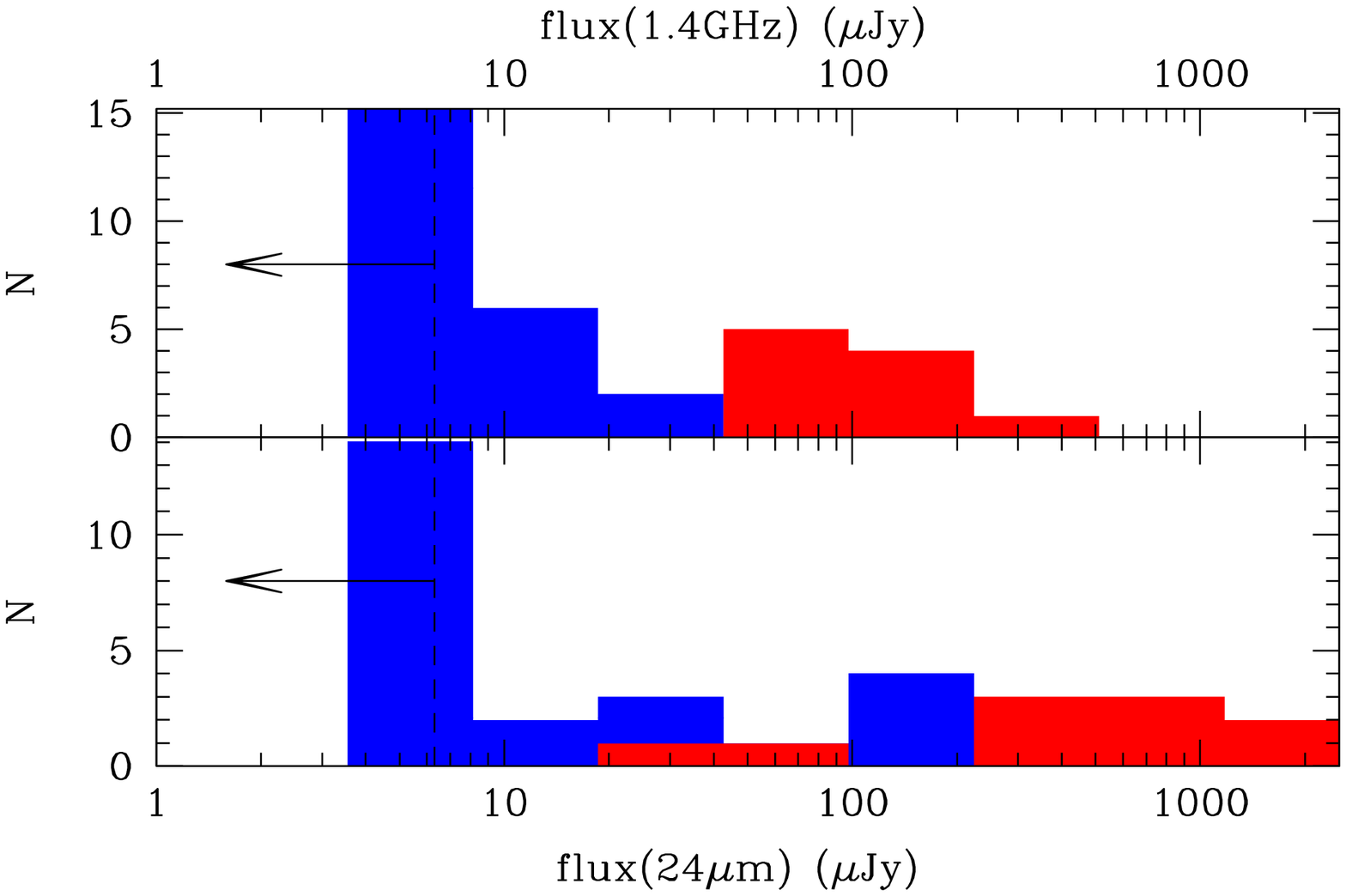}
\caption{\footnotesize
{\bf Upper windows:} 
Comparison of the  Reddy et al.\ (2008) infrared (IR) luminosity function for $z=1.9-2.7$ (translated into SFR using Kennicutt 1998) to the SFRs for SMGs (red) derived from 24$\mu$m, and for UV-selected galaxies derived from
exponentially declining star formation models (blue).
The IR luminosity function of just the UV-selected galaxies in GOODS-N is also shown (not corrected for
completeness) as squares, for direct comparison to the spike galaxies (which also are not corrected for completeness at the low luminosity end). As expected, the vast majority of the overdensity compared to the average field appears at the high-SFR end in SMGs.
Below is shown the raw distributions in SFR for UV-galaxies and SMGs/SFRGs.
{\bf Lower windows:} 
For the z$\sim1.99$ spike galaxies, distributions in radio (1.4~GHz) flux and 24$\mu$m flux density are presented, to show consistent measures across UV-selected and SMG/SFRG populations, unaffected by uncertainties in SFRs calculated by different methods. }
%These raw data emphasize that the UV-inferred SFRs for BX galaxies are often overestimated }
\label{fig3}
\vskip0.5cm
\end{figure}

\subsection{Significance of the structure}

There is an obvious $z\sim1.99$ galaxy concentration in the SMG redshift distribution 
lying  in the GOODS-N field which coincides with an apparently less significant concentration in the UV-selected
galaxies.  Figure~1 shows the angular distribution of the galaxies, revealing that the 9 SMGs and SFRGs in a redshift {\it spike} are not obviously clustered spatially across the field shown in Figure~1, spanning a region 7\arcmin\ (3.5~Mpc) on a side, although there are two relatively close pairs (of 9\arcsec\ and 17\arcsec, 76~kpc and 143~kpc respectively).
The redshift distributions around the $z\sim2$ spike are compared in Figure~2.

We first aim to characterize this structure with respect to galaxies selected at both UV and submm/radio wavelengths.
%We first define and discuss the selection functions in both the SMGs and UV-selected galaxies, and then go on to estimate the level of contrast. 
%Finally, we relate the redshift space contrast to matter overdensities.
%
For the SMG sample, we define a selection function by smoothing the redshift distribution of the $\sim$100 SMGs in the expanded survey from C05 described above. 
We adopt a cluster finding algorithm based on an Erlangian distribution function (Eadie et al.\ 1971)
\begin{equation}
p(\Delta z\,|\,N \lambda) = \lambda (\lambda\,\Delta z)^{N-2} \exp(-\lambda\,\Delta z) / (N-2)!
\end{equation}
which characterizes the probability (in the absence of clustering) that a group
of $N$ galaxies would span a redshift interval $\Delta z$, considering that the product of
our selection function and the density of SMGs is $\lambda$.
The procedure is very similar to that described in Steidel et al.\ (1998), and was used for our studies of halo substructures in the Andromeda galaxy (e.g., Chapman et al.\ 2006). 

\begin{figure}
\centering
\includegraphics[width=7.9cm,angle=0]{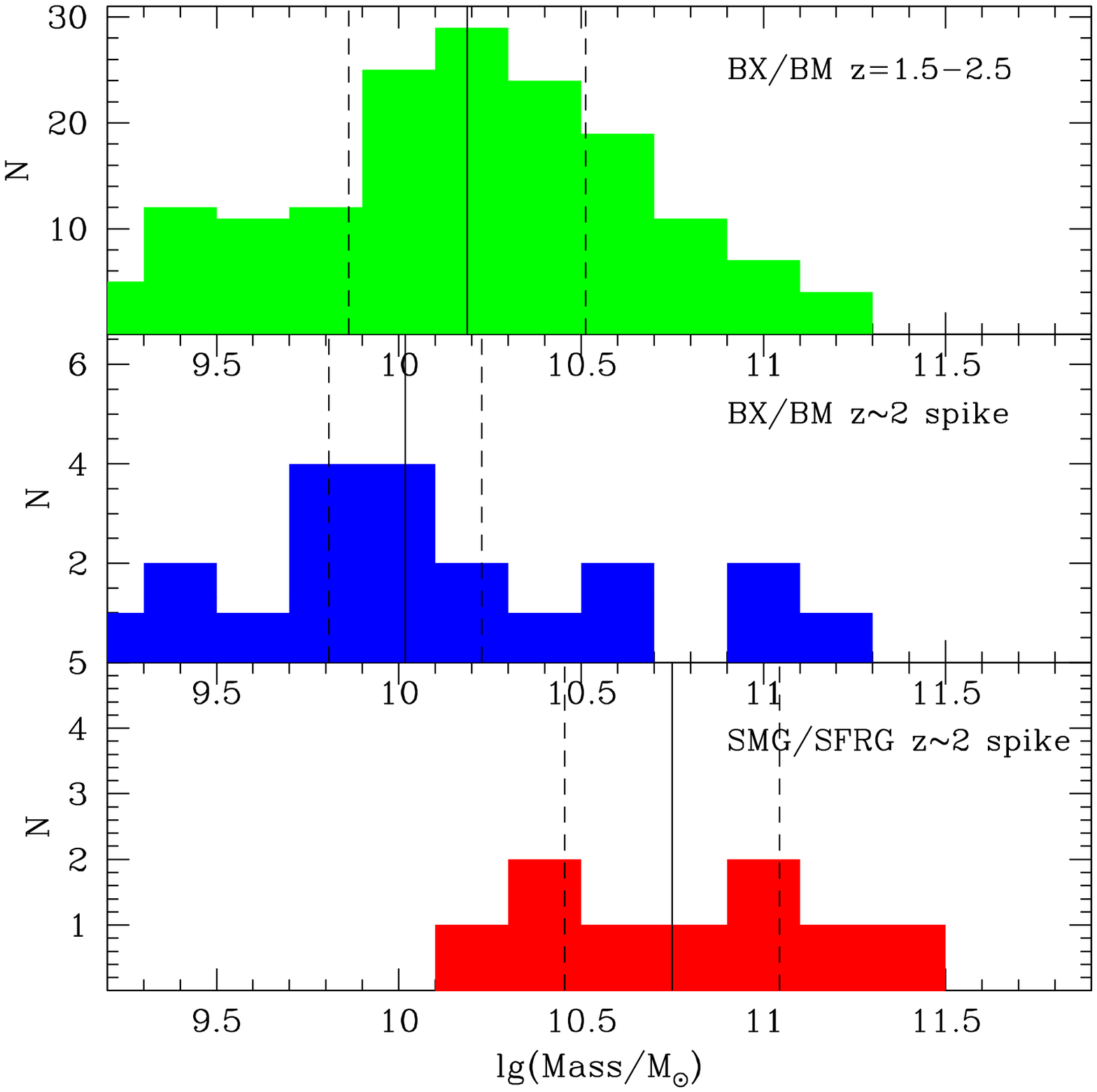}
\vskip-5cm
\includegraphics[width=7.9cm,angle=0]{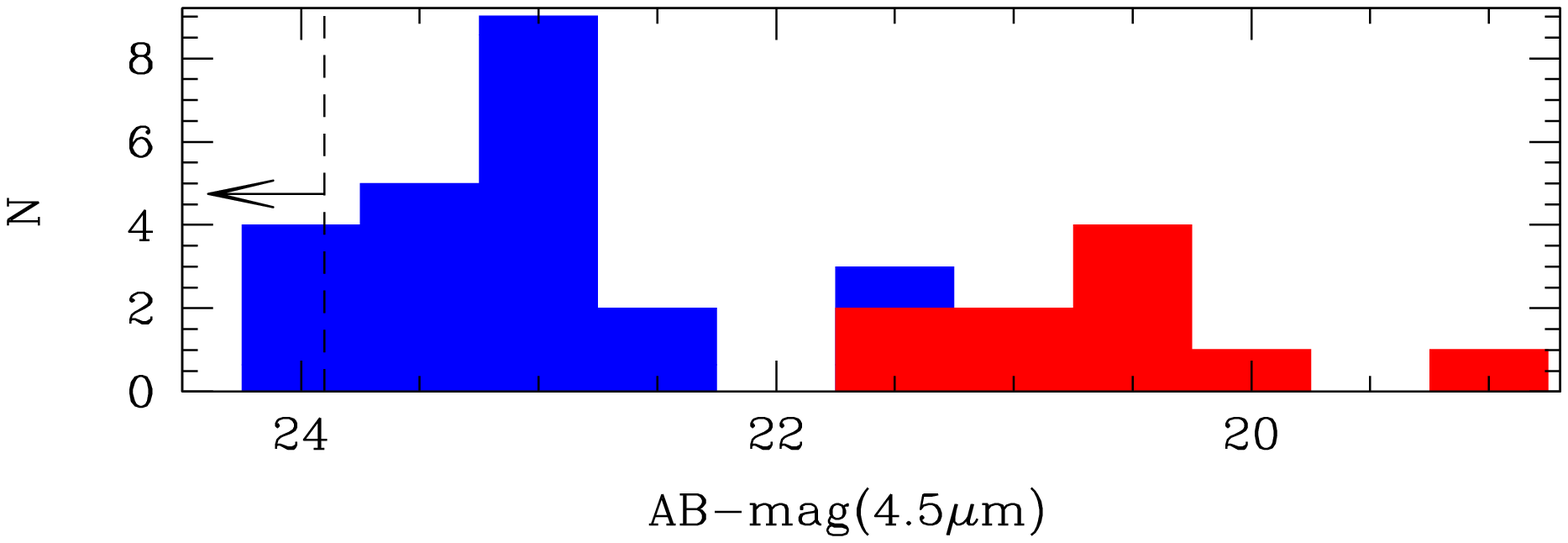}
\caption{\footnotesize
{\bf Top windows:} Comparison of histograms of stellar masses for SMGs in the spike, and for UV galaxies derived from
 continuous star formation models of Reddy et al.\ (2006), both in and out of the spike. The median values and bootstrap errors are shown with lines. Neither the UV galaxies nor the SMGs show increased stellar mass buildup compared with the field, although the shear presence of the 9 SMG/SFRGs in the spike suggests a higher volume averaged stellar mass at $z=1.99$ than in the field.
 {\bf Bottom window:} For the z$\sim1.99$ spike galaxies, distributions in 4.5$\mu$m AB-mag (IRAC channel 2), representing a rest-frame $\sim$1.6$\mu$m or $H$-band light, showing the same trend as the stellar mass calculated above from the complete SEDs.}
\vskip0.5cm
\label{fig5}
\end{figure}

First looking just at the six SMGs in the structure, we find a
probability of 0.005\% for the  concentration of SMGs at $<z>=1.992$ with $\Delta z=0.008$ to be drawn from the selection function by chance, which maximizes the contrast with a peak of 5 of the 6 SMGs lying within the region. We define a galaxy overdensity as the number of galaxies found in the structure (excluding the two galaxies which define the edges of the redshift interval) divided by 
the expected number in this redshift interval from the selection function (which is is only 0.4), yielding 
$\delta_g^z=4/0.4 = 10$.
%{\bf the 1200~km/s region centered on their median redshift - ACTUALLY it is just z=1.99, dz=0.08} . 
This result is obviously dependent on the precise redshift  adopted, as some scatter is introduced when individual galaxies are defined by different combinations of lines from molecular, neutral and
ionized gas, which can  differ by as much as 1000~km/s in a given SMG  (Chapman et al.\ 2005, Greve et al.\ 2005, Swinbank et al.\ 2004). 
If the SMGs and SFRGs are considered together, the contrast is maximized with 7 of the 9 galaxies, and the probability to find the structure by chance rises marginally to 0.007\%, with an implied overdensity $\delta_g^z=8$.
%This peak corresponds to $\delta\equiv n_{\rm cluster} / \bar n - 1 = 6.0^{+1.6}_{-1.2}$ assuming Poisson uncertainties.

Second,  for the UV-selected galaxies, we adopt the selection function defined by the smoothed distribution of all galaxies in the spectroscopic survey of Steidel et al.\ (2004) and Reddy et al.\ (2006): see Fig.~1. 
%is it worth the discussion on the selection function being 
The probability of the $z\sim1.99$ concentration being drawn at random from the selection function is 11.3\%, and this contrast is maximized with a peak of 15 galaxies in the $<z>=1.987$, $\Delta z=0.015$ region, with an overdensity $\delta_g^z=n_z/(n-1) = 2.5$.
While C05 showed that many SMGs do have the colours of UV-selected galaxies, these  spike SMGs and SFRGs  do not overlap at all with 
the  UV-selected galaxies in this $z\sim1.99$ structure.
%We highlight a non-overlapping sample of galaxies zoomed in around the $z\sim2$ redshift spike in Fig.~1b.
%
%The SMGs comprise 7 galaxies in the structure, whereas the UV-selected galaxies encompass 24 galaxies total in a plausible Gaussian structure centered on the median redshift of the 7 SMGs.

We have therefore identified a highly significant overdensity in ultra-luminous SMGs which appears not to correspond to a similarly overdense structure in less luminous, UV-selected galaxies. 
The UV-galaxies  have much lower SFRs
and longer lifetimes (Reddy et al. 2006) than SMGs, and it is a reasonable hypothesis that the UV-population would represent a more dependable tracer of large-scale structure. 
While there remains the possibility that redshift incompleteness in quiescent galaxy populations (for example) could be responsible for the discrepancy, we will proceed with our explicit findings of a different bias between the two star forming galaxy populations. 

\subsection{Analysis of the $z=1.99$ spike galaxies}

The environmental context of high-redshift galaxy clusters is a topic which has lately received interest, since the preferential buildup of stellar mass in clusters should happen at earlier epochs than for field galaxies. % ... SFRs.
Steidel et al.\ (2005) found considerable excess stellar mass buildup and longer mean stellar ages in the  densest  structure from their $z\sim2$ surveys, as  compared with the coeval field population.
Clustering results from large spectroscopic surveys (e.g., VVDS --Meneux et al.\ 2008), rest-$K$ mass-selected surveys using {\it Spitzer}-IRAC (e.g., de La Torre et al.\ 2007), and red galaxy selection
(e.g,. BzK/ERO galaxies -- Daddi et al.\ 2003) have come to similar conclusions regarding these environmental dependencies.
Our naive explanation of the differential bias between the submm and UV-selected galaxies invokes a modest mass cluster in an active formation period, and we might therefore expect to find little in the way of stellar mass or age differences in the spike population. However, our interpretation could be clouded by missing populations of evolved galaxies which are not well sampled by GOODS-N spectroscopic surveys. We therefore analyze the spike galaxies to see if they show environmental effects characteristic of other high-$z$ clusters.

First, we look at SFRs for all the galaxies in the spike (Fig.~3). Since many of the UV-selected galaxies are not detected in the 24$\mu$m or radio bands, while the SMGs and SFRGs typically have their SFRs underestimated by large factors in the UV, we calculate SFRs for the UV-galaxies from the dust-corrected UV luminosity, and from the 24$\mu$m luminosity for the SMGs and SFRGs. For the spike galaxies we also show the distributions of radio (1.4~GHz) and 24$\mu$m flux density to demonstrate that our derived SFRs are consistent with the raw data.
The clustered environment allows for a useful comparison of SFR indicators without K-correction and model galaxy uncertainties.
Our new radio measurements for the spike galaxies detect significant numbers of the lower luminosity UV-selected galaxies in addition to all the SMGs and SFRGs, and radio-derived SFRs are generally consistent with those derived from {\it Spitzer} 24$\mu$m fluxes.
%Table~1 reveals that these flux densities are reasonably correlated.
%, COMPARE SFR-uv to radio, 24$\mu$m.
%
We then consider the luminosity function (LF) of the whole $z\sim1.99$ spike
%, taking the 24$\mu$m luminosity  as a reliable estimate of SFR, 
and compare with the LF  derived for the whole UV population by Reddy et al.\ (2006) in Fig.~3. As expected for a relatively insignificant structure in the UV, the Reddy et al.\ LF is reasonably consistent with the $z\sim2$ field galaxies, with a small excess expected from the
overdensity factor of order two. 
When added to the spike sample, the SMGs clearly form a significant excess  compared with the average LF of all (dust-corrected) UV-selected galaxies.

% ZZZ - justifies the UV tracing of LSS that was  alluded to in the abstract, but I
%got a vague sense that this was touched on in the text before. Maybe  this discussion would be
%better part of setting the context rather than the beginning of he  results?

%ZZZ - Is this a bit of a speculation - maybe there are lots of EROs hanging  
%about, and we're looking at something that's got a few fueling monsters left.

Next, we consider the stellar masses of galaxies in the spike (Fig.~4), compared to all non-spike UV-selected galaxies.
The UV-galaxies in the structure do not exhibit the enhanced buildup of stellar mass which is apparent 
in other high-$z$ UV/optical-selected clusters (Steidel et al.\ 2005, Kurk et al.\ 2008). The UV spike galaxies have $<$M$^*$$>=1.1\pm3.6\times10^{10}$~M$_\odot$, while the ``field'' UV galaxies have
$<$M$^*$$>=1.5\pm3.4\times10^{10}$~M$_\odot$ (derived from the tables in Reddy et al.\ 2006). 
Using the stellar masses estimated for the SMGs by Hainline (2008), Hainline et al.\ (in prep.), we find that the spike SMGs clearly have larger stellar masses ($<$M$^*$$>=6.0\pm3.6\times10^{10}$~M$_\odot$) than the UV-selected galaxies, %in or out of the spike, 
but the SMGs' stellar masses are
not significantly different from the median for 73 SMGs from Hainline (2008) 
(M$^*=6.9\times10^{10}$~M$_\odot$). We note that Borys et al.\ (2005) derived even larger ($5\times$) stellar masses for these spike SMGs. 
Hainline (2008) ascribe the difference in their masses to those from
Borys et al.\ as due to two issues. Firstly  the use of different stellar
evolution models: the Maraston TP-AGB models (Maraston et al.\ 2006), and differing light-to-mass
and SF histories, and secondly the removal of AGN contributions to
the observed 8$\mu$m flux prior to fitting.  Alexander et al.\ (2008) have also
explicitly attempted to correct for the AGN contribution to the rest-frame
K-band light in these SMGs and derive lower masses (factor of $\sim2$).  We
stress that the relative masses of the BX/BMs and SMGs will be robust
against some of the systematic uncertainties in the models, and that our
results are not sensitive to changing the masses by factors of a few, and
hence we believe our conclusions are robust.
We also show in Figure~4 the 4.5$\mu$m fluxes of the spike galaxies (rest-frame $\sim1.6\mu$m  or $H$-band) revealing the SMGs are clearly much brighter than the UV-galaxies without uncertainties in stellar mass model fitting.
There is therefore evidence for significant stellar  
buildup in our $z\sim1.99$ overdense environment, although this is driven exclusively by the large overdensity of SMGs.

Comparing our ``spike'' SMGs to``typical"  SMGs at similar epochs (e.g., Chapman et al.\ 2005, Borys et al.\ 2005, Pope et al.\ 2006, Hainline et al.\ 2008)  reveals no  obvious differences in SFR or stellar mass  (the median values of both are similar for spike and field SMGs -- $<$SFR$>$=810~M$_\odot$/yr, $<$M$^*>$=7$\times10^{10}$~M$_\odot$).  An additional route to studying the activity in the structure versus the field  would be through AGN populations. Cataloged AGN in GOODS-N (e.g., Cowie et al.\ 2005, Reddy et al.\ 2007) do not reveal any significant enhancement at $z\sim2$, nor do the Xray properties of these spike SMGs (Alexander et al.\ 2005).

%{\bf IAN:  
%any evidence of enhanced AGN activity in the spike? -- hard to quantify relative to any other samples}

% *all* SMGs might   be in spikes (of some description) if you had large/complete enough  surveys.

%Chandra extended Xray source centered on Casey et al.\ (2008) OFRG giant elliptical.

\section{Discussion}

\subsection{Evolutionary context of the structure}

To place our SMG cluster in a cosmological context, we must assume a galaxy bias for SMGs.
We adopt $b\sim 2.1$, a value consistent with simulations in van Kampen et al.\ (2005), and similar to the  
estimated galaxy bias in the UV galaxy sample (Adelberger et al. 2005b).
Following the method introduced by Steidel et al.\ (1998), we relate the true mass overdensity
$\delta$  to the observed redshift-space
overdensity of SMGs $\delta_g^z$ through the expression 
\begin{equation}
1+b\delta=\left|C\right|(1+\delta_g^z)
\end{equation}
where $\left|C\right|\equiv V_{app}/V_{true}$ takes into account the effects of 
redshift-space distortions from peculiar velocities. %, which is itself a function of both $\delta$ and $z$. 
For our SMGs, SFRGs and UV galaxies, $V_{app} \simeq 9900$ Mpc$^{3}$ is 
the co-moving volume in which the
measurement is being made, bounded by the 10$'$0 by
10$'$0 region on the plane of the sky and the co-moving distance, 
neglecting peculiar velocities, between $z=1.978$ and $z=2.008$. 
The volume correction factor can be approximated %(see Steidel et al.\ 1998)
by $C=1+f-f(1+\delta)^{1/3}$ where $f=\Omega_m(z)^{0.6}=0.96$  
at $z \simeq 2.0$. 
Thus $\delta\simeq 2.3$
with $\delta_g^z=10.0$, and $C=0.53$. 
The linear overdensity,  $\delta_0$, can then be approximated using equation 18 (spherical collapse) 
from Mo \& White (1996)
$$\delta_0\simeq A_1(1+\delta)^{-2/3}+A_2(1+\delta)^{-0.5866}+A_3(1+\delta)^{-1/2}+A_4$$
with $A_1=-1.35$, $A_2=0.78785$, $A_3=-1.12431$, $A_4=1.68647$.
Thus for our $z\sim1.99$ SMG structure  ($\delta=2.3$), we find $\delta_0\simeq0.85$
as the real space linear matter overdensity.

We now address what the SMG overdensity will have become
by redshift $z\sim 0$.  Evolved forward in our adopted cosmology, its linear over-density of $\delta_0\sim 0.85$ at $z=2$
corresponds to a linear over-density of $\delta_p\sim 2.13$ at $z=0$.
This exceeds the collapse threshold $\delta_c=1.69$. % in the SMG case.

%ZZZ - Define b after eq 1 - it's referenced later.

However, if we repeat the analysis with the UV-selected galaxy $\delta_{UV}^z\sim2.5$ case,
we find a matter overdensity at $z=2$ of only $\delta=0.83$, a linear overdensity $\delta_0=0.5$ or $\delta_p\sim 1.1$ at $z=0$, and the structure associated with this apparently modest overdensity of UV-galaxies would not have collapsed by $z=0$.  

The mass of a virialized
object, as we would infer for the SMG overdensity, can be calculated by assuming  that the SMG overdensity is associated with an Eulerian matter
overdensity of $\delta_m \simeq 2.3$ at $z=2.0$ (e.g., Steidel et al.\ 1998), so its mass is
$(1+\delta_m)\bar\rho V_{true}\sim10^{15} M_\odot$
where $\bar\rho$ is the mean co-moving matter density and 
$V_{true}=V_{app}/C \sim 18700$ Mpc$^3$ is approximately the volume we calculated
within the observed overdensity.
There is, however, no straightforward way to estimate the mass of the lower contrast UV-galaxy spike, as it has not obviously virialized by $z=0$. We simplistically assume the inferred mass would be $>$3$\times$ lower.

%SUBSEC: toy model for bias
\subsection{Explaining the differential bias}

Our task is then to explain why such a significant structure in the SMG population exhibits such a modest  overdensity 
in the UV-selected galaxies.
One possibility is that the true overdensity and cluster mass concentration is in fact as large as implied by the SMGs, and we have missed a large population of evolved quiescent galaxies, selected for instance 
through the restframe 2$\mu$m emission. 
%ZZZ - Would the GOODS-N photo z catalog not show up a possible z~2 red sequence?
At a redshift $z\sim2$, such a population is difficult to confirm 
spectroscopically, and might explain for instance why we found no difference in the stellar masses of field versus spike UV-selected star forming galaxies. 
However, we see no logical reason why the mass concentration should then be represented by SMGs (and not UV-galaxies), which are an even more luminous population of star forming galaxies.

Another line of inquiry is to look for mechanisms to produce the apparent differential bias directly.
%Some variant on 
Dynamical masses have been estimated for galaxies selected in the restframe UV using nebular emission line widths (Pettini et al.\ 2000; Erb et al.\ 2006), and for SMGs and SFRGs using both
nebular lines and rotational CO molecular lines 
(e.g., Frayer et al.\ 1998; Neri et al.\ 2003; Swinbank et al.\ 2004, 2006, 2008; Chapman et al.\ 2008).
Attempts to account for the fact that estimates of the dynamical masses of high-$z$ galaxies are significantly less than the clustering-infered halo masses (Adelberger et al.\ 2005b; Blain et al.\ 2004) 
have fallen under the names of 
temporal bias (Scannapieco \& Thacker), merger bias 
(e.g., Furlanetto \& Kamionkowski 2006) or assembly bias (Dalal et al.\ 2008). 
These approaches attempt to account for the rapid mergers and other properties of galaxies in halos boosting the apparent clustering signal, 
and they might naturally extend to the scenario we have uncovered in this $z\sim2$ structure,
which represents an even more radical version of this problem.
As pointed out by Furlanetto \& Kamionkowski (2006), the extended Press-Schechter and Mo \& White (1996) biasing scheme yields no merger bias as it explicitly ignores the variation of merger rates with the large-scale density field.
However, for populations like SMGs, even an analytic approximation to the bias implicit in aggressively
merging galaxies is suggestive that the clustering length could be enhanced.
The challenge is to apply this formalism not just to a pair-wise enhancement of the population, but to the
synchronization of 9 SMG-class galaxies spread over an angular region 3.5~Mpc on a side.

We consider from Furlanetto \& Kamionkowski (2006), equation 24, 
which describes how  
pairs of peaks in the density field are clustered, as a function of  
the clustering of the peaks themselves (i.e., identifying the peaks with  
galaxies, and the ``pairs" as recently merged objects).  In the limit  
that the peaks are well above the characteristic mass scale of objects  
(likely a reasonable assumption for SMGs), the correlation function of pairs $\chi(r)$ 
is

\begin{equation}
\chi(r) = [ 1 + \xi_g(r)]^4 - 1
\end{equation}

where $\xi_g(r)$ is the correlation function of the galaxy population of  
interest ($b_g$ is the associated matter correlation function).  This gives an  
enhancement to the overall clustering of the underlying galaxy  population.

Ignoring redshift space corrections, we then make the following  
argument:  there is one true matter overdensity, $\delta_m$.  Suppose  
that UV-galaxies and SMGs are drawn from halos with the same mass (and hence  
bias $b_{UV}$).  Then, if SMGs correspond to pairs of objects, they have  
an extra ``pair" bias $b_p = \sqrt(\chi/\xi_g)$.  So the observed  
overdensities $\delta^z$ will be:

\begin{equation}
1 + \delta^z_{UV} = 3.5 = 1 + b_{UV} \delta_m
\end{equation}
\begin{equation}
1 + \delta^z_{SMG} = 11 = 1 + b_p b_{UV} \delta_m
\end{equation}
which requires $b_p\sim4$ (slightly different if 
redshift space correction is included).

While the model is approximate and the observations are not necessarily complete, the
result is not unreasonable.  A remaining question is  
what is the correct $r$ which applies in eq.\ 2\&3 above to evaluate $b_p$.   
On very large scales, where $\xi_g<<1$, we can Taylor expand eq.\ 3 about $r=0$,
$\chi(r) \sim 4 \xi_g(r)$,
to  infer a pair bias $b_p \simeq 2$.
Our physical interpretation of this is two pairs (or four SMGs), so we obtain four  
factors of $\xi_g$.  
However, on smaller scales, the pair bias goes up because  
one is dealing with peaks:  on scales where the correlation function of SMGs is large, $\xi_g\sim1$ (i.e., where  
clusters are forming), $\chi(r)=15$, so $b_p\sim4$.  This is similar to our required factor of boosting, and likely sufficient given uncertainties in the observations.

We emphasize that simulations that have tried to  
quantify this ``assembly bias" give smaller effects, $\sim10$\%  (Wechsler et  
al.\ 2002; Zao et al.\ 2003a,b; Dalal et al.\ 2008). However, these studies haven't looked at this exact question of recent mergers, which  
may be a stronger effect than the concentration, age, spin, etc., which are modeled in these simulations,
and therefore our toy model is not inconsistent with these studies.

\subsection{Context and Implications for SCUBA2 surveys}

While the most overdense
forming environments at $z>1.5$ have been observed to have large numbers of ultra-luminous galaxies associated (e.g., the SSA22 $z=3.09$ proto-cluster has 8 known SMGs -- Geach et al.\ 2006),
our results suggest that if one
searches for galaxy clusters using the most vigorously bursting (and typically highest luminosity) systems, one is prone to finding the brief periods when systems of lower matter overdensity are rapidly evolving.
The exceptionally strong concentration of SMGs presented here corresponds to a modest, not statistically significant (in terms of $\delta$) concentration in UV-selected galaxies,
whereas the expectation would be that this submm ``spike'' should represent one of  the largest mass structures known.
While we have outlined a toy model to explain the scenario in \S~3.2, it remains to be seen how such scenarios can be realized by detailed theoretical models and simulations of cluster formation.

%- maybe there are lots of EROs hanging  
%about, and we're looking at something that's got a few fueling monsters left.

%%% negrello stuff and beams/submm clusters

%%% baugh/sommerville/swinbank studies of luminous galaxies in context

%%% millenium no merger bias
Baugh et al.\ (2005) have presented a model of the SMG population in a semi-analytical galaxy evolution framework. Using the the Millennium Simulation (Springel et al.\ 2005), Almeida et al. (2008 in prep.) predict that the correlation function of model SMGs from Baugh et al.\ (2005) is close to a power law 
($\eta = (r/r_o)^\gamma$) over more than three decades in separation, with 
$\gamma=-1.94\pm0.05$ and a correlation length $r_0=8.8\pm0.3$~Mpc, slightly smaller but consistent with that measured by Blain et al.\ (2004) from the redshifts of SMGs, 
$r_0 = 9.8\pm3.0$~Mpc.
Swinbank et al.\ (2008) performed a detailed comparison of actual SMG observations with model SMGs in the Baugh et al.\ (2005) model.
In the Millennium Simulation, this correlation length corresponds to a predicted halo mass of 
M$_{\rm halo} = 3.1^{+5.7}_{-1.9}\times10^{12}$~M$_\odot$ (Gao et al. 2005). 
Measuring the dark halo mass for the model SMGs directly, Swinbank et al.\ (2008) derive 
M$_{\rm halo} = 3.6^{+5.5}_{-1.5}\times10^{12}$~M$_\odot$. 
Taken together this suggests there  is no merger bias in the correlation lengths: the correlation  
length of the SMGs is identical, within the errors, with the whole  population of halos of
the same masses.
Our discovery in this paper of a large submm burst in a modest mass environment suggests that 
 a more thorough  investigation is needed of the clustering properties of SMGs using  
combined semi-analytic and numerical simulations to trace the star  
formation and merger histories of high-redshift galaxies.
%even in the Baugh et al.\ (2005) semi-analytic treatment of burst-mode star formation in mergers, the  variation of merger rates with the large-scale density field is not sufficiently well represented.

%with the halo masses measured in the model SMGs directly, we therefore Þnd no 
%strong evidence for a merger bias in the measurements for the observed galaxy clustering.

We note that similar criticisms could apply to the clustering of AGN (e.g., Gilli et al.\ 2005), and overdensities selected through AGN (e.g., Rottgering et al.\ 1996, 2003; Brand et al.\ 2003).
In particular, the overdensities of other galaxies around luminous AGNs are typically poorly characterized relative to the field. Our well calibrated results on SMGs suggest that these AGN-identified galaxy clusters could be far less massive than suggested.

While our characterization of a single cluster of SMGs cannot be extrapolated to cluster formation in general, it is noteworthy that large scale surveys in the submm with SCUBA2, and the {\it Herschel} and {\it Planck} space
telescopes, and even existing wide-field surveys at 24$\mu$m from {\it Spitzer} (e.g., Farrah et al.\ 2006) may reveal rather modest mass 
structures associated with some of the highest contrast source overdensities.
The large beam size of {\it Planck} and the {\it Herschel}-Spectral and Photometric Imaging REceiver (SPIRE) 500$\mu$m survey may be particularly sensitive to this sort of effect (Negrello et al.\ 2005, 2007), where the low-resolution surveys will measure the summed contributions of groups of sources within the beam.
Further, the number of these low-mass but {\it active} clusters are likely significant enough to
make it impossible to associate a large $r_0$ with a large dark matter halo mass. 
Detailed followup of the highest concentrations of ultra-luminous sources, although arduous, is clearly warranted.
In the future, serendipitous searches for CO emitters at the same  
redshift using ALMA, and evidence for boosted flux from confused sources using  
{\it Herschel}
may also provide insight into the nature of this structure.

%{\bf Some
% speculation about how the SMGs know it's time to burst over 1Mpc?}

%\subsection{Characterization of other redshift spikes in the SMG/OFRG surveys, and implications for general SMG cross-correlation with UV-selected galaxies}

%Our proposal, that the SMGs can occur preferentially in short but very 
%active periods of modest mass clusters, suggests the Adelberger et al.\ (2005)
%concern about cluster statistics for SMGs is not significant, while at least partially  solving the apparent %paradox in Blain et al.\ (2004) that too many SMG associations exist to be progenitors of relatively 
%rich clusters at the present epoch.

%We proceed first to characterize other overdensities in the GOODS-N field, and relate them to other overdensities identified in the Blain et al.\ (2004) analysis of SMGs

%\vskip2cm
\section{Conclusions}
%
%How do the 24um inferred SFRs for SMGs+SFRGs differ from the UV-selected galaxies (which are completely orthogonal to the SMGs+SFRGs in this structure).
%
%Discussion of the SA22 $z=3.09$ result, and possible larger extent of the cluster. Could we have somehow just found a low density edge of a super-cluster?
%Are we finding a limb of a filamentary structure where cold flows stimulate more SMGs etc.?
%
We have described an apparent cluster of SMGs and RGs at $<z>=1.99$ in the GOODS-N field, the strongest known association of SMGs (Blain et al.\ 2004; Chapman et al.\ 2005).
The implied matter over-density from the redshift space contrast $\delta^z_g\simeq10$ is 
is $\sim10^{15}$~M$_\odot$.
Searching the same volume in UV-selected galaxies, we find a much reduced overdensity
 ( $\delta^z_g\simeq2.5$). 
% and the implied dark matter halo mass scale is $<$3$\times$ less.
The low SFRs of UV-galaxies and much longer star formation timescales on average compared with the SMGs and SFRGs suggest they should be more dependable for defining the mass scale of this structure.
We conclude that we are observing a modest mass overdensity
(which likely will not even virialize by $z=0$), which is undergoing a particularly
active star formation phase.
%(to be explained in some heuristic manner, since it isn't an obvious byproduct of any simulations)

We have characterized the bolometric luminosity function of the $z\sim1.99$ spike galaxies compared to that estimated for a large sample of UV-selected star forming galaxies (Reddy et al.\ 2008), finding that it is almost entirely in the SMG population where an excess is seen. The clustered environment allows for a useful comparison of SFR and stellar mass indicators without K-correction and model galaxy uncertainties.
Our new radio measurements for the spike galaxies detect significant numbers of the lower luminosity UV-selected galaxies in addition to all the SMGs and SFRGs, and radio-derived SFRs are generally consistent with those derived from {\it Spitzer} 24$\mu$m fluxes.

%However, timescale of the luminosity must be key. 
While the most overdense
forming environments at $z>1.5$ have been observed to have large numbers of ultra-luminous galaxies associated, 
%(e.g., the SSA22 $z=3.09$ proto-cluster has 8 known SMGs -- Geach et al.\ 2006),
our results here suggest that if one
searches for galaxy clusters using the highest luminosity systems, one is prone to
finding the brief periods when systems of lower matter overdensity are rapidly evolving.
SMGs don't obviously trace the most massive structures in the Universe, as simulations and constraints to date have suggested they should. 
The success of our toy model with merger bias explaining the basic elements of our observations  suggests that simulations may not currently be accurately reflecting the processes occurring in the formative phases of galaxy clusters. Further work to embed these aspects of merger bias in simulations will therefore be of great interest in understanding galaxy cluster assembly.
Investigation of similar systems in upcoming surveys at long wavelengths will be of great interest to explore the full range of environments associated with strong overdensities of ultra-luminous galaxies.

\section*{acknowledgements}

We thank Steve Furlanetto for very helpful discussions on extending his work on merger bias to our SMG cluster, as well as the anonymous referee who for their suggestions.
SCC acknowledges an NSERC Discovery grant and a fellowship from the Canadian Space Agency 
which supported some of this work.
IRS acknowledges support from the Royal Society.

\begin{table*}
\begin{center}
\caption{Properties of all UV-selected galaxies, SMGs and SFRGs in GOODS-N $z=1.99$ spike$^a$} 
\label{tableSat}
\begin{tabular}{lllcccccl}
\hline
{ID} & {RA} & dec & redshift$^b$ & type of redshift  & S$_{\rm 1.4 GHz}$\,$^c$ &  S$_{\rm 24 \mu m}$ & 
M$_{\rm AB, 5.8 \mu m}$ & M$_{\rm AB, 4.5 \mu m}$ \\% & other names$^a$ \\
\hline
SMG-93  &  12 36 00.13 & +62 10 47.2  & 1.994 & UV, H$\alpha$ & 128.5$\pm$8.1 (26.5) & 1237$\pm$8.0 & 20.32$\pm$0.27 & 20.97$\pm$0.08  \\
SMG-132 & 12 36 18.32 & +62 15 50.5  & 1.999 & CO(4-3) & 172.0$\pm$8.4 (34.3) & 344.0$\pm$8.0 & 20.31$\pm$0.18 & 20.65$\pm$0.07 \\ %GN06   \\
SMG-140e & 12 36 21.25 & +62 17 08.3 & 1.994 & Spitzer & 169.4$\pm$8.8 (31.9)& 370.0$\pm$11.4  & 20.06$\pm$0.10 & 20.31$\pm$0.06 \\% GN07\\
SMG-140w &12 36 20.99 & +62 17 09.5 &1.989/1.988 & UV,H$\alpha$ & 63.4$\pm$10.6 (9.0) & {...}$^d$ & 20.67$\pm$0.09 & 20.96$\pm$0.07\\%GAL 24.15 0.17 0.26 0.08 0.59 0.09 22.82 0.21 21.75 0.19 21.19 0.07 20.96 0.07 20.67 0.09 21.00 0.14 263.7 16.4
SMG-169 & 12 36 32.53 & +62 07 59.7 & 1.993 & UV & 80.4$\pm$8.6 (15.1) & 820.8$\pm$15.2 & 19.64$\pm$0.15 & 20.29$\pm$0.07  \\
SMG-172$^e$ & 12 36 35.57 & +62 14 24.0 & 2.001 & UV & 77.0$\pm$7.8 (16.3) & 1410.0$\pm$13.9 & 18.29$\pm$0.09& 18.90$\pm$0.06 \\ 
SMG-255 & 12 37 11.99 & +62 13 25.6 & 1.992 & UV, CO & 50.2$\pm$8.1 (10.2) & 225.0$\pm$7.0 & 21.79$\pm$0.21 & 21.45$\pm$0.09 \\ %{GN39} \\
\hline
SFRG-130 & 12 36 17.54 & +62 15 40.7 & 1.993 & UV & 234.3$\pm$8.6 (45.1) & 20.2$\pm$5.2 & 20.77$\pm$0.17 & 20.58$\pm$0.08 \\
SFRG-179 & 12 36 40.73 & +62 10 11.0 & 1.968/1.977 & UV & 72.5$\pm$8.3 (13.7) & 83.0$\pm$7.8 & 21.29$\pm$0.19 & 21.33$\pm$0.07\\
SFRG-254 & 12 37 11.32 & +62 13 30.9 & 1.996/1.993 & H$\alpha$/CO(4-3) & 127.4$\pm$8.7 (23.0) & 537.3$\pm$9.1 & 19.72$\pm$0.11 & 20.01$\pm$0.06\\ %AGN/SB components \\
\hline
BX-1201 & 12 36 14.13 &+62 15 41.8 & {...}/2.000 & UV & 10.6$\pm$7.0 (2.8) & 29.0$\pm$5.5 & 23.04$\pm$0.27 & 22.99$\pm$0.07 \\%GAL 24.00 0.17 0.18 0.07 0.71 0.11 999 999 999 999 22.94 0.07 22.99 0.07 23.04 0.27 999 999 29.0 5.5
BX-1192 & 12 36 16.83 &+62 15 14.3 & {...}/1.996 & UV & 21.3$\pm$10.3 (3.1) & 111.6$\pm$10.7 & 21.47$\pm$0.15 & 21.56$\pm$0.07\\%GAL 24.22 0.17 0.15 0.07 0.86 0.11 999 999 999 999 21.52 0.07 21.56 0.07 21.47 0.15 21.67 0.22 111.6 10.7
BX-1071 & 12 36 20.10 &+62 11 12.6 & {...}/1.996 & UV & $<31.8$& {...} & {...} & 23.06$\pm$0.21\\%GAL 24.41 0.17 0.27 0.08 0.76 0.16 23.71 0.28 24.10 0.49 22.84 0.10 23.06 0.21 999 999 999 999 999 999
BX-1228 &12 36 20.19 &+62 15 40.6 &1.999/1.995 & UV & $<14.1$ & 14.7$\pm$4.1 & 23.05$\pm$0.22 & 23.10$\pm$0.11\\%GAL 24.03 0.17 0.39 0.08 0.65 0.09 23.12 0.27 23.33 0.30 22.94 0.08 23.10 0.11 23.05 0.22 999 999 14.7 4.1
BX-1267 &12 36 22.67 &+62 16 21.6 &1.996/1.996 & UV & $<14.5$ & 48.7$\pm$7.2 & {...} & {...}\\%GAL 23.90 0.14 0.13 0.05 0.51 0.09 22.84 0.19 22.17 0.18 23.45 0.30 999 999 999 999 999 999 48.7 7.2
BX-1339 &12 36 25.09 &+62 17 56.8 &1.993/1.984 & UV & $<14.9$ & {...} & 22.63$\pm$0.31 & 23.03$\pm$0.12 \\%GAL 24.60 0.20 -0.05 0.08 0.48 0.07 999 999 999 999 23.06 0.09 23.03 0.12 22.63 0.31 999 999 999 999 #huge wind!
BX-1307 &12 36 48.33 &+62 14 16.7 &{...}/2.002 & UV & 18.2$\pm$6.4 (5.2) & 145.5$\pm$12.2 & 21.34$\pm$0.09 &21.61$\pm$0.07\\%GAL 23.30 0.11 0.20 0.05 0.74 0.07 22.69 0.17 21.83 0.14 21.69 0.07 21.61 0.07 21.34 0.09 21.44 0.26 145.5 12.2
BX-1329 & 12 36 54.62 &+62 14 07.7 & {...}/1.987 & UV & $<18.2$& {...} & {...} & {...}\\%GAL 24.69 0.20 -0.04 0.08 0.45 0.07 23.95 0.34 23.79 0.35 24.46 0.25 999 999 999 999 999 999 8.2 2.5
BX-1827 &12 36 56.63 &+62 15 19.0 &1.988/{...} & UV & 9.6$\pm$4.1 (2.3)& {...} & {...} &23.46$\pm$0.13\\%GAL 24.84 0.20 0.56 0.10 1.12 0.20 24.48 0.32 23.28 0.32 23.42 0.09 23.46 0.13 999 999 999 999 999 999
BX1129 &12 36 56.94 &+62 08 48.7 &{...}/1.973 & UV & 21.7$\pm$11.2 (3.0)& 104.4$\pm$10.3 & 21.46$\pm$0.09 &21.60$\pm$0.07\\%GAL 22.80 0.13 0.21 0.04 0.58 0.05 999 999 999 999 21.52 0.07 21.60 0.07 21.46 0.09 22.00 0.13 104.4 10.3  1SIGMA WING
BX-1431 &12 36 58.48 &+62 16 45.5 &2.006/1.996 & UV & $<13.4$& {...} & {...} &23.25$\pm$0.34\\%GAL 24.00 0.17 0.09 0.07 0.45 0.06 >25.0 999 23.13 0.29 23.11 0.16 23.25 0.34 999 999 999 999 999 999
BX1378 &12 37 02.02 &+62 14 43.4 &{...}/1.971 & UV & 10.4$\pm$4.2 (2.5)& 29.4$\pm$5.5 & 22.43$\pm$0.10 &22.40 0.07\\%GAL 23.90 0.14 0.33 0.06 0.66 0.09 23.14 0.22 22.75 0.23 22.42 0.07 22.40 0.07 22.43 0.10 23.15 0.18 29.4 5.5 1SIGMA WING
BM-1122 &12 37 02.62 &+62 11 56.7 &1.994/1.986 & UV & $<17.6$ & {...} & {...} & 23.30$\pm$0.16\\%GAL 23.96 0.14 0.17 0.05 0.33 0.06 24.18 0.50 24.00 0.46 23.27 0.10 23.30 0.16 999 999 999 999 999 999 #huge wind
BX-1434 &12 37 16.80 &+62 14 38.8 &{...}/1.994 & UV & 4.6$\pm$6.1 (1.5) & {...} & {...} &23.18$\pm$0.27\\%GAL 24.49 0.17 0.26 0.08 0.51 0.11 23.62 0.27 23.21 0.28 23.14 0.19 23.18 0.27 999 999 999 999 999 999
BM-69 &12 37 17.68 &+62 14 35.9  &1.991/1.991 & UV & $<17.8$ & 32.7$\pm$5.8 & {...} & {...}\\%GAL 25.12 0.18 0.31 0.10 0.27 0.07 >25.0 999 >24.4 999 999 999 999 999 999 999 999 999 32.7 5.8
BX-184 &12 37 19.28 &+62 13 00.6 &1.998/{...} & UV & $<12.2$ & {...} & {...} &22.95$\pm$0.25\\%GAL 24.22 0.17 0.70 0.12 0.99 0.15 23.42 0.28 23.46 0.30 22.80 0.07 22.95 0.25 999 999 999 999 999 999
BM-70 &12 37 20.05 &+62 14 57.1 &1.997/1.994 & UV & $<30.9$ & {...} & {...} &23.69$\pm$0.19\\%GAL 24.05 0.17 0.37 0.08 0.15 0.05 23.11 0.27 23.33 0.30 23.24 0.07 23.69 0.19 999 999 999 999 999 999
BX-1605 &12 37 21.51 &+62 18 30.6 &1.977/1.970 & UV & $<23.1$& {...} & {...} &23.36$\pm$0.17\\%GAL 23.89 0.14 -0.07 0.05 0.19 0.04 999 999 999 999 23.29 0.07 23.36 0.17 999 999 999 999 999 999 1SIGMA WING
BM-1155 &12 37 23.98 &+62 12 12.1 &2.024/2.015 & UV & $<33.6$ & {...} & {...} &23.28$\pm$0.13\\%GAL 23.99 0.14 0.08 0.05 0.15 0.05 23.25 0.22 23.50 0.25 23.21 0.07 23.28 0.13 999 999 999 999 999 999 1SIGMA WING
BX-1642 &12 37 32.40 &+62 17 50.8 &2.010/2.004 & UV & $<21.0$ & {...} & {...} &22.95$\pm$0.09\\%GAL 24.29 0.17 0.25 0.08 0.46 0.07 999 999 999 999 22.97 0.09 22.95 0.09 999 999 999 999 999 999 #huge wind!
BX-1708 &12 37 32.65 &+62 19 10.6 &{...}/1.987 & UV & 16.3$\pm$9.1 (3.0) & 208.3$\pm$14.6 & 21.03$\pm$0.07 &21.23$\pm$0.07\\%GAL 24.50 0.20 0.45 0.10 1.00 0.15 999 999 999 999 21.46 0.07 21.23 0.07 21.03 0.07 21.34 0.08 208.3 14.6
BX-1694 &12 37 33.82 &+62 18 46.3 &2.009/2.005 & UV & 10.1$\pm$8.4 (2.0) & 13.8$\pm$4.3 & 22.28$\pm$0.11 & 22.58$\pm$0.07\\%GAL 23.66 0.14 0.11 0.05 0.53 0.07 999 999 999 999 22.50 0.07 22.58 0.07 22.28 0.11 999 999 13.8 4.3
\hline
SCBX-6106 &12 37 10.82 & +62 08 19.0 &{...}/1.984 & UV & $<13.5$ & {...} & 22.41$\pm$0.55 &23.15$\pm$0.22\\
SCBX-6946 &12 36 57.70  & +62 10 36.8  &{...}/1.992 & UV & $<14.7$ & {...} & 22.53$\pm$0.31 & 22.84$\pm$0.25\\
%SCBX-7758 &12 37 05.76 & +62 13 03.2 &1.997/1.994 & UV & $<12.3$ & {...} & 10 &\\ steidel z=0.01
%SCBX-5952 &12          38       36.86         &+62           7       49.01&1.997/1.994 & UV & 100 & 24 & 10 &\\
\hline
%spike $>1\sigma$ wings & & & & & & & \\%surrounding
\hline
\end{tabular}
\end{center}
{\footnotesize
\begin{tabular}{l}
%$^a$ GN naming comes from  et al.\ 2006, 2008.\\
$^a$ Galaxies listed have redshifts which lie within $\pm2\sigma$ of the $z=1.99$ peak, and as shown by the vertical bars in Fig.2b.\\
$^b$ Redshifts of UV-selected galaxies are quoted as {\it emission line/absorption line}, where {...} indicates that one or the other was not\\ possible from the spectrum -- see Reddy et al.\ (2006) for details of the BX/BM galaxies, whereas we have adopted this same procedure\\ for the SCBX galaxies (from our own redshift surveys), selected from a color-cut transformed to AB-mag from that presented in\\ Steidel et al.\ (2004).
Molecular gas (CO) redshifts are from Chapman et al.\ (2008), and Smail et al.\ (in prep). The
{\it Spitzer-IRS}\\ redshift comes from Pope et al.\ (2008).\\
$^c$ Radio fluxes from Morrison et al.\ (2008), derived from the full resolution (1.7\arcsec) map. Integrated flux measurements and errors are\\ shown, with the signal-to-noise in brackets.  For UV-galaxies, source fluxes are listed if a radio peak lies within 0.5\arcsec\ of the optical\\ source, otherwise a 3$\sigma$ upper limit.\\
$^d$ 24$\mu$m flux density of SMG-H140w is confused with SMG-H140e, although the radio emission is clearly resolved by MERLIN\\ (Chapman et al.\ 2004), and even by the VLA (Biggs \& Ivison 2006).\\
$^e$ HDF172 could conceivably be interpreted as an SFRG rather than an SMG. It is a SCUBA photometry detection in Chapman et al.\\ (2001), a SCUBA-mapping non-detection in Borys et al.\ (2004), and a weak ($\sim3\sigma$) MAMBO-mapping source (at 1.2mm) from\\ Greve et al.\ (2008).\\
\end{tabular}}
\vskip0.5cm
\end{table*}

%\begin{table*}
%\begin{center}
%\caption{Properties of UV-selected in $z=1.99$ spike} %, and $z=1.87$, $z=2.09$ spikes}
%\label{tableSat}
%\begin{tabular}{lllccccccc}
%\hline
%{id} & {RA} & dec & redshift & S$_{\rm 1.4GHz}$ & S$_{\rm 850\mu m}$ &  $R_{AB}$ mag & radio morph & optical morph & spec class \cr
%\hline
%BXX  & 12:36:17.57  &   62:15:40.7 & 1.993 &  200.0 12.8 & 0 & 24.02  &-1.16 1.08 \\
%\hline
%\end{tabular}
%\end{center}
%{\footnotesize
%\begin{tabular}{l}
%$^a$ The radio source extends over 3.3\arcsec\,
%$^a$ 
%\end{tabular}}
%\end{table*}

\end{document}